# Cardiac CT segmentation based on distance regularized level set

Xin Yang Wu [1] (✉)

[1] Sun Yat-sen University, Shenzhen, China
wuxy97@mail2.sysu.edu.cn

**Abstract.** Before analyzing the CT image, it is very important to segment the heart image, and the left ventricular (LV) inner and outer membrane segmentation is one of the most important contents. However, manual segmentation is tedious and time-consuming. In order to facilitate doctors to focus on high-tech tasks such as disease analysis and diagnosis, it is crucial to develop a fast and accurate segmentation method [1]. In view of this phenomenon, this paper uses distance regularized level set (DRLSE) to explore the segmentation effect of epicardium and endocardium [2], which includes a distance regularized term and an external energy term. Finally, five CT images are used to verify the proposed method, and image quality evaluation indexes such as dice score and Hausdorff distance are used to evaluate the segmentation effect. The results showed that the method could separate the inner and outer membrane very well (endocardium dice = 0.9253, Hausdorff = 7.8740; epicardium Hausdorff = 0.9687, Hausdorff = 6. 3246).

## 1      Introduction

Cardiovascular disease (CVD) is one of the most common causes of death in the world [3]. X-ray computed tomography (CT) cardiac imaging can provide detailed anatomical information such as chambers, blood vessels, coronary arteries and coronary artery calcification score, thus greatly deepening people's understanding of the anatomical structure of cardiovascular diseases. The analysis and diagnosis are based on the heart image segmentation, and the LV segmentation is very important.
Although level set function (LSF) has been widely used in medical image segmentation [4], the evolution process of traditional level set has been plagued by irregularity, which leads to numerical error and eventually destroys the stability of level set evolution. In order to overcome this difficulty, people use the numerical method of reinitialization to restore the regularity of LSF and maintain the stable level set evolution [5]. However, this method may mistakenly move the zero-level set away from the expected position.

   In order to avoid the problem of reinitialization, I choose the distance regularized level set model. The general segmentation process is that the image is preprocessed by bilateral filtering, and then the initial ellipse template is framed in the intima. In the iterative process, the parameters are constantly modified to get a perfect fit of intima segmentation, and then the outer membrane segmentation is obtained by expanding.

The structure of this paper is as follows: the second section introduces the segmentation algorithm, including the principle of distance regularization level set and parameter setting; The third section introduces the experimental results, through the visual effect and statistical indicators to prove the feasibility of this method. After that, the settings of parameters are discussed.

## 2 Method

The overall framework is shown in the figure below.

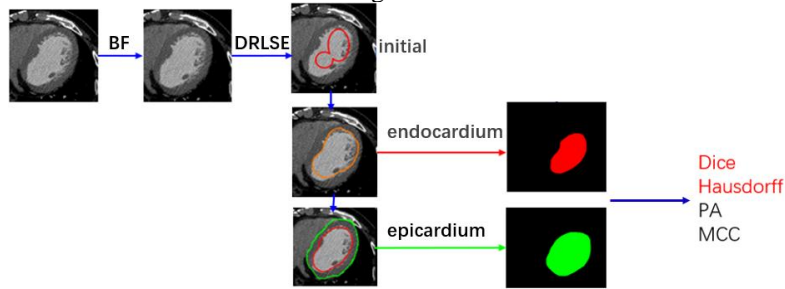

**Fig 1.** the basic structure of the algorithm

### 2.1 Bilateral Filtering

Because the level set segmentation directly using the original image is affected by noise, and the computational cost is huge [6], it is necessary to preprocess the image. After trying a variety of image enhancement methods, the bilateral filtering method is finally determined because of its good effect in removing CT spatial noise [7]. The main purpose of this step is to enhance the edge structure and remove noise in the middle, so that the level set is less affected by the noise in the non edge area, quickly updated, and accurately iterated and did not leak when approaching the boundary.
Bilateral filtering is nonlinear. The principle is to give different weights according to the distance and brightness of the target pixels to achieve the denoising effect. The larger the intensity difference and the farther the distance between the points around the target pixel, the smaller the weight. The formula is as follows:

$$y[k] = \frac{\sum_{n=-N}^{N} W[k,n] \cdot x[k-n]}{\sum_{n=-N}^{N} W[k,n]} \quad (1)$$

The weight W includes the spatial information $W_S$ and the intensity information $W_i$. The farther the two target pixels are, the larger the value difference and the smaller the weight is. In this way, the surrounding noise is filtered out because of the large intensity difference, and the edge structure is retained because of its close distance and heavy weight.
## 2.2 Distance regularized level set (DRLSE) segmentation

### 2.2.1 Level Set Function
This method is based on the traditional level set function(LSF). LSF is a technique of using fixed mesh to represent moving interface or boundary. Based on an implicit surface, it can distinguish whether each point is on the zero level line by whether the function value is equal to zero. The level set equation can be obtained by deriving and simplifying the level set definition and adding the numerical stability term on the right side

$$\frac{\partial \Phi}{\partial t} + \mathbf{u} \cdot \nabla \Phi = \gamma \nabla \cdot \left( \varepsilon \nabla \Phi - (1-\Phi) \frac{\nabla \Phi}{|\nabla \Phi|} \right) \quad (2)$$

Where ε represent the width of the transition band in ϕ from 0 to 1, which is a constant in the given domain, and it is equal to the maximum value of the grid size h in the domain. Parameter γ determines the amount of reinitialization or stabilization of level set functions. Through this equation, the free surface can update the zero profile in fluid motion according to the fluid velocity [8].

### 2.2.2 The Distance regularized level set model
In order to keep the good state of LSF and make the evolution of level set stable and accurate, it is necessary to keep it smooth during the evolution process. Therefore, DRLSE introduces the energy formula with distance regularization. It is defined by

$$\varepsilon(\phi) = \mu R_p(\phi) + \lambda L_g(\phi) + \alpha A_g(\phi) \quad (3)$$

where $R_P(\Phi)$ is the regularization term, $\lambda$ is the weight of the weighted length term in the energy regularization formula, and $\alpha$ is the weight of the weighted area term. Then the DRLSE equation can be expressed as

$$\frac{\partial \phi}{\partial t} = \mu \, div \left( d_p(|\nabla \phi|) \nabla \phi \right) - \frac{\partial \varepsilon_{ext}}{\partial \phi} \quad (4)$$

The edge indicating function g is defined as follows:

$$g \triangleq \frac{1}{1 + \nabla G_\sigma * I^2} \quad (5)$$

$G_\sigma$ is the Gaussian kernel with the standard deviation σ, convoluted with image I to smooth noise. The value of this function at the boundary of the object is usually less than that at other locations.

## 2.3 Evaluating Indicator

The results were evaluated by two dimensions of area and distance, which were dice score and Hausdorff distance. Because the last two indicators were similar to dice score, their weights were lower.

### 2.3.1 Dice Score

It is a commonly used indicator for image segmentation. The larger the value is, the more similar the segmentation is

$$Dice = \frac{2 * A_{sl}}{A_s + A_l} \quad (6)$$

where $A_s$ is the area of the segmentation result area, $A_l$ is the area of the label, $A_{sl}$ is the area where the two overlap.

**2.3.2 Hausdorff Distance**

The similarity is described by the distance between two point sets. The smaller the value, the better. The distance is defined by

$$d_H(X,Y) = \max\left\{\sup_{x \in X}\inf_{y \in Y} d(x,y), \sup_{y \in Y}\inf_{x \in X} d(x,y)\right\} \quad (7)$$

where d (x, y) represents the distance between two points, X and Y represent the segmentation result and label respectively.

**2.3.3 Pixel Accuracy (PA)**

The ratio represents the proportion of the number of correctly classified pixels to the total number of pixels.

**2.3.4 Matthews Correlation Coefficient(MCC)**

MCC is essentially the correlation coefficient between observed and predicted binary classification that returns to a value between -1 and +1.

$$MCC = \frac{TP \times TN - FP \times FN}{\sqrt{(TP+FP)(TP+FN)(TN+FP)(TN+FN)}} \quad (8)$$

## 3    Results

Five ventricular CT images were used in this experiment, and the resolution of each image was 726×726. After exploring the influence of different enhancement methods, border selection and iterative methods on the results, the optimal method is determined.

**3.1    Segmentation Results**

The following shows the image segmentation results of five images, the index statistics of CT Figure 1, and the comparison between this method and the traditional LSF method.

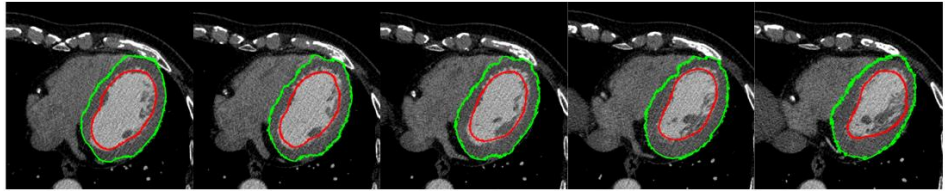

**Fig 2.** The red border is the inner membrane and the green border is the outer membrane.

**Table 1.** Statistics of internal and external membrane indexes.

|  | Dice | Hausdorff | PA | MCC | Dice |
|---|---|---|---|---|---|
| Endocardium | 0.9253 | 7.8740 | 0.9840 | 0.9166 | 0.9253 |
| epicardium | 0.9687 | 6.3246 | 0.9888 | 0.9620 | 0.9687 |

**Table 2**. Comparison of inner and outer film segmentation effect between DRLSE and LSF.

|  |  | Dice | Hausdorff | PA | MCC |
|---|---|---|---|---|---|
| Endocardium | LSF | 0.8967 | 8.4857 | 0.8996 | 0.8965 |
|  | DRLSE | 0.9253 | 7.8740 | 0.9840 | 0.9166 |
| epicardium | LSF | 0.9001 | 7.0841 | 0.9236 | 0.8968 |
|  | DRLSE | 0.9687 | 6.3246 | 0.9888 | 0.9620 |

### 3.2. The Influence of Enhancement Methods

In order to improve the computational efficiency and accuracy, it is necessary to enhance the image before segmentation. In the ablation experiment, I explored the combination effect of ① bilateral filtering(BF), ② Wiener filtering (WF)[9], and ③ gradient operator [10], and evaluated the results of rough endometrial segmentation with dice and Hausdorff as the main indexes when the other parameters were consistent in 85 iterations.

**Table 3**. The effect of different pre-treatment on the segmentation of intima.

|  | Dice | Hausdorff | PA | MCC |
|---|---|---|---|---|
| BF+gradient | 0.8984 | 7.6811 | 0.9787 | 0.8865 |
| WF+gradient | 0.8939 | 6.7082 | 0.9765 | 0.8823 |
| BF | 0.9003 | 7.0711 | 0.9776 | 0.8902 |
| WF | 0.8735 | 7.7460 | 0.9701 | 0.8635 |
| Gradient | 0.8975 | 7.8740 | 0.9787 | 0.8857 |
| Original | 0.9003 | 7.4833 | 0.9785 | 0.8897 |

### 3.3. The Influence of Initial Border

The ablation experiment studied the influence of the shape and position of the border on the results.

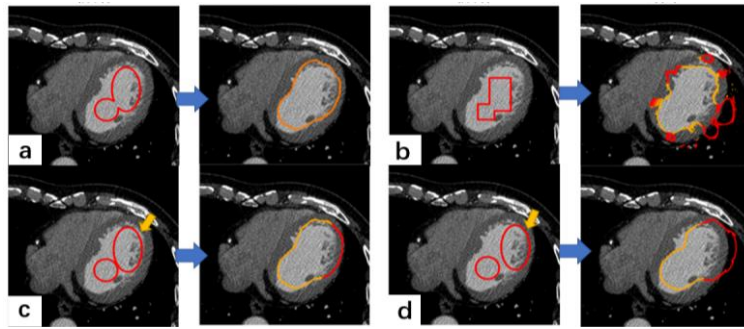

<p style="text-align:center">**Fig 3.** Analysis of initial border and segmentation results.</p>

### 3.4. The Influence of Iterative Method

In the ablation experiment, I also studied the effects of iterative methods (1) from the inside out and (2) from the outside to the epicardium.

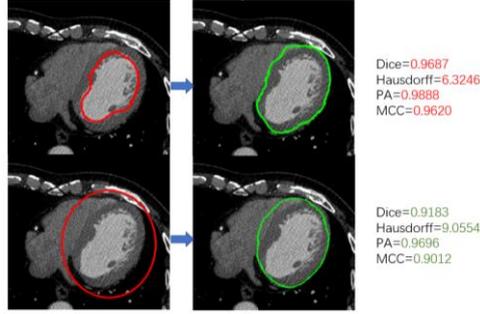

**Fig 4.** different outer membrane iteration methods and results. In the figure above, the inner membrane is used as the initial segmentation to iterate outwards, while in the figure below, the inner membrane is iterated outwards.

## 4 Discussion

This algorithm involves $\alpha$, $\varepsilon$, $\mu$, $\sigma$, $\lambda$, and the number of iterations. To sum up, the key to accurate segmentation is to find and approach the critical value ε as much as possible with the continuous reduction of $\alpha$. Although larger ε will produce jagged lines on the boundary, but it will not cause huge error after smoothing the image every 10 iterations.

However, there is still room for improvement in practice. First of all, although the results can be improved by using bilateral filtering enhancement, the filter is essentially a single resolution, which is unable to access different frequency components of the image. Although it is very efficient to remove noise in the high-frequency region, the performance of removing noise in the low-frequency region is poor [7]. As a result, the segmentation in the irregular inner and outer film boundaries cannot be optimized very well, such as spots near the inner membrane.

On the other hand, although DRLSE speeds up the whole process, it still needs a large amount of computation, since the input image size was large (726 × 726) and due to system configuration, the iteration speed was still very slow, up to 1 minute, which brings great trouble to debug parameters.

Finally, the DRLSE algorithm is too sensitive to initialize. The segmentation result is closely related to the initial image and parameter setting. For each image, the parameters and initialization must be set separately, so it is not universal.

## 5 Conclusion

This paper explored the effectiveness of the level set LV segmentation through

DRLSE model. After exploring various enhancement filtering, initial contour and iterative methods, by comparing the dice score, Hausdorff distance and other indicators of segmentation results and the visual effect of iterative process, the best segmentation method was determined. After bilateral filtering, the inner membrane was segmented iteratively by using the ellipse template, and then the outer membrane was segmented by using the inner membrane results.